\documentclass[prl,showpacs,superscriptaddress,twocolumn]{revtex4}
\usepackage{graphicx}
\usepackage{dcolumn}

\newcommand{\beq}{\begin{equation}}
\newcommand{\eeq}{\end{equation}}
\newcommand{\beqa}{\begin{eqnarray}}
\newcommand{\eeqa}{\end{eqnarray}}

\renewcommand{\vec}[1]{{\mathbf{#1}}}

\newcommand{\dagga}{{\phantom{\dagger}}}

\begin{document}

\title{The $T=0$ heavy fermion quantum critical point
as an orbital selective Mott transition}

\author{Lorenzo \surname{De Leo}}
\affiliation{Centre de Physique Th\'{e}orique, Ecole Polytechnique,
CNRS, 91128 Palaiseau, France}
\affiliation{Department of Physics, Rutgers University,
Piscataway, NJ 08854, USA}

\author{Marcello Civelli}
\affiliation{Theory Group, Institut Laue Langevin, 38042 Grenoble Cedex, France}

\author{Gabriel Kotliar}
\affiliation{Department of Physics, Rutgers University,
Piscataway, NJ 08854, USA}


\begin{abstract}

We describe the $T=0$ quantum phase transition in heavy fermion
systems as an orbital selective Mott transition (OSMT) using a
cluster extension of dynamical mean field theory. This transition
is characterized by the emergence of a new intermediate energy
scale corresponding to the opening of a pseudogap and the
vanishing of the low-energy hybridization between light and heavy
electrons. We identify the fingerprint of Mott physics in heavy
electron systems with the appearance of surfaces in momentum space
where the self-energy diverges and we derive experimental
consequences of this scenario for photoemission, compressibility,
optical conductivity, susceptibility and specific heat.

\end{abstract}

\pacs{71.27.+a, 71.10.Hf, 75.20.Hr, 75.30.Mb}

\maketitle

Heavy fermion materials containing electrons in open $4f$ or $5f$
shells and in broad $spd$ bands, continue to be a subject of
great interest in condensed matter physics
\cite{rmp07-heavyfermions}. The description of the
antiferromagnetic-paramagnetic transition in these systems is
highly non trivial, because, in addition to the fluctuations of
the magnetic order parameter, one has to take into account
the changing character, from itinerant to localized, of the $f$
electrons\cite{johansson,lanata}.
Recent publications\cite{pepin07,demedici05} have debated whether
the quantum phase transitions observed in heavy fermions can be
described as an orbital selective Mott transition (OSMT), i.e. a
Mott transition taking place in the $f$ orbitals with the
$spd$ orbitals remaining itinerant.
Some finite temperature aspects of this phenomena are captured by
{\it single site} dynamical mean field theory
(DMFT)\cite{rmp96,dmft-heavy,demedici05}, a method that
captures well the peculiar dynamics of the Mott transition
assuming a purely local approximation. In
Refs.\cite{Paul07,pepin07,pepin08} however, the authors have
shown by means of slave boson techniques that going beyond the
local approximation is an essential ingredient to obtain an OSMT
at zero temperature.

In this paper we overcome the spatial limitation of DMFT by using
one of its cluster extensions (the cellular DMFT,
CDMFT\cite{kotliar01}), which takes into account short-ranged
correlation, and we demonstrate the existence of a $T=0$ OSMT. In
Ref.\cite{deLeo07}, we have fully characterized the phase diagram
of a heavy fermion model across a quantum critical phase
transition, separating a strongly renormalized Fermi liquid from
an antiferromagnetic phase. Here, by constraining the mean-field
non-ordered solution, we focus on the qualitative evolution of
the electronic structure. In this way, we isolate the physics
that stems directly from the localization of the $f$ electrons
from the physics of the magnetic order that intervenes at low
temperature in a given material.

We show in particular that at the transition a new energy scale
emerges. In this energy range a pseudogap opens
in the $f$ spectra and the hybridization between heavy $f$ and light
$spd$ electrons goes to zero, leading to a complete decoupling
of the two bands.
Beyond this energy range the $f$-$spd$ hybridization remains finite.
These phenomena have a clear interpretation
in terms of an OSMT, revealed by the appearance of
surfaces of diverging self-energy in
momentum space, fingerprint of a Mott
mechanism (mottness\cite{stanescu07}).
In the conclusions, we derive a set of experimental consequences
relevant for the normal state of real materials close
to the quantum critical point at temperatures above
the ordered state.

We study the quantum phase transition driven by a hybridizing
parameter $V$ in the periodic Anderson model, which describes
free $spd$ electrons locally hybridized to non dispersing
strongly correlated $f$ electrons. The Hamiltonian is:
\begin{eqnarray}\label{Hamiltonian}
    H \!\!&=& \! \sum_{\vec{k}} ( \varepsilon_\vec{k} -\mu)
    d^\dagger_{\vec{k}\sigma} d^\dagga_{\vec{k}\sigma}
    + V \sum_\vec{k} \left( f^\dagger_{\vec{k}\sigma}
    d^\dagga_{\vec{k}\sigma} + h.c. \right) \nonumber\\
    &&+ (E_f-\mu) \sum_\vec{k} f^\dagger_{\vec{k}\sigma}
    f^\dagga_{\vec{k}\sigma}
    + U \sum_i f^\dagger_{i\uparrow}f_{i\uparrow} f^\dagger_{i\downarrow}f_{i\downarrow}
\end{eqnarray}
where $d^\dagger_{\vec{k}\sigma}$ [$f^\dagger_{\vec{k}\sigma}$]
creates an $spd$ [$f$] electron with momentum $\vec{k}$ and spin
$\sigma$.
The conduction band dispersion is
$\varepsilon_\vec{k} = -1/3(\cos k_x + \cos k_y + \cos k_z)$, the
other parameters $U=10$, $\mu=0.2$ and $E_f-\mu=-5.7$. The $spd$
and $f$ electrons Green's functions can be written in terms of
the $f$ electron self-energy $\Sigma$:
\begin{equation}\label{Gf}
    G_{\alpha}(\omega,\vec{k})=\, \left[ \, \omega+ X_{\alpha}(\vec{k},\omega)
    -\frac{V^{2}}{\omega+ Y_{\alpha}(\vec{k},\omega)} \,\right]^{-1}
\end{equation}
where $\alpha=f, spd$, $X_{f}(\vec{k},\omega)=
\mu-E_{f}-\Sigma(\vec{k},\omega)$, $Y_{f}(\vec{k},\omega)= \mu-
\varepsilon_{\vec{k}}$ and $X_{spd}= Y_{f}$, $Y_{spd}= X_{f}$.

We implement CDMFT on a two-site
cluster\cite{deLeo07}. We believe this is the
minimal unit able to capture the physics close to
the transition point. The Hamiltonian in Eq.~(\ref{Hamiltonian}) is
mapped onto an effective two impurity Anderson model (2IAM) and
solved self-consistently via the Lanczos method\cite{caffarel94},
which introduces a finite energy resolution on the Matsubara axis\cite{rmp06}
$\omega_n= (2n-1)\pi/\beta$, with $\beta=100$.

In order to physically interpret our results, we extract the
momentum dependent lattice self-energy $\Sigma(\vec{k},\omega)$
in Eq.~(\ref{Gf}) from the cluster quantities, restoring the cubic
lattice symmetry. Various methods have been proposed in the
literature\cite{rmp06}. Close
to the transition, where particles tend to localize, it has been
shown\cite{stanescu06} that a suitable quantity to adopt is the
cluster cumulant $ \hat{M}(\omega)=\, \left[ \left( \omega + \mu-
E_{f}
   \right) \hat{\mathbf{1}} - \hat{\Sigma}(\omega)
   \right]^{-1}
$. In our case we have
$M(\vec{k})= M_{0}+ \frac{1}{3} \, M_{1} ( \cos k_{x}+ \cos
k_{y}+ \cos k_{z} )$ where $M_0=M_{11}=M_{22}$ and
$M_1=M_{12}=M_{21}$, and
$\Sigma(\vec{k},\omega)=\, \omega+ \mu- E_{f}- M(\vec{k},\omega)^{-1}$.
A stringent self-consistent test of this
periodization can be obtained by re-calculating the
local $f$ electron Green's function $\sum_{\vec{k}} G_{f}(\vec{k},\omega)$,
via Eq.~(\ref{Gf}), and confronting it with the cluster counterpart,
direct output of the CDMFT calculation.
\begin{figure}[!tb]
\begin{center}
\includegraphics[width=8cm,angle=-0]{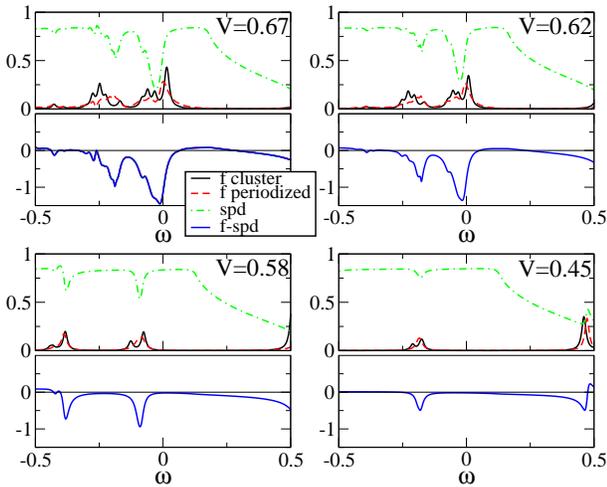}
\caption{The local density of states (DOS) as a function
of the hybridization $V$ in a low energy window around
the chemical potential (Hubbard bands are out of the picture at $\omega\sim \pm 5$). Black
continuous lines are the cluster $f$-electron DOS.
The red dashed [green dot-dashed] lines are the
$f$[$spd$]-electron DOS $-\frac{1}{\pi}\sum_{\vec{k}} \, G_{f
[spd] }(\vec{k},\omega)$.
The bottom panels show the effective hybridization
$\sum_\vec{k} \hbox{Re}G_{f-spd}(\omega,\vec{k})
$ (blue continuous line).} \label{spectra}
\end{center}
\end{figure}
In Fig. \ref{spectra} we show the low-energy imaginary parts of the local
$f$ Green's functions (the density of states DOS). The good
agreement between the periodized $f$ DOS
and the cluster $f$ DOS validates
our procedure. Moreover we show the DOS
for the $spd$ electrons
and the effective hybridization $\hbox{Re}
G_{f-spd} =  \sum_\vec{k} \hbox{Re} G_{f-spd}(\omega,\vec{k})
$\cite{rmp96}. These quantities
demonstrate that, as a function of the tuning parameter $V$, the
system undergoes a phase transition.
The numerical uncertainties become greater near the transition,
hence we cannot determine whether the transition is second order,
as predicted in a scaling theory\cite{Continentino}
(in which case the best fit of the pseudogap scaling with a
power law gives an exponent $z\nu \sim 0.33$),
or first order as found in a recent Guzwiller
treatment\cite{lanata}.
For $V>V^*\sim 0.58$ the
system is in the heavy-fermion phase where the $f$ electrons
present a Kondo peak at the Fermi level $\omega=0$ and take active part in
the conduction. The strong hybridization with the $spd$ electrons
is evident in the suppression of the $spd$ DOS and in the
non-zero value of the effective hybridization $f-spd$ close to $\omega=0$.
The intensity of the $f$ peak reduces approaching
$V^*$, while at the same time the $spd$ spectral weight enhances.
For $V<V^*$ the system is in an orbital selective Mott state
where the $f$ electron spectrum has a gap.
The $f$ electron spectral weight is not completely transferred from
low energy to the Hubbard bands (placed around $\omega\sim \pm5$) but
rather to a new intermediate energy scale, giving rise to a
pseudogap. Within this pseudogap the $spd$
electrons recover the free band DOS and the
effective hybridization is zero. This shows that the $spd$ band
at low energy is completely decoupled from the $f$ band, but the
effective hybridization remains active at a finite intermediate
energy scale.

We can now display the quasiparticle bands along some specific cuts
in the $\omega-\vec{k}$ space. From Eq.~(\ref{Gf}), we notice that
$G_{f}$ transforms into $G_{spd}$ upon the exchange of
$E_{f}-\Sigma(\vec{k},\omega)$ with $\varepsilon_{\vec{k}}$.
The poles of the $f$ and $spd$ Green's
function are therefore the same,
provided the Im$\Sigma_{\vec{k}}$ is small as in
Fermi liquid theory, and 
the same resolving equation is obtained for either $\alpha=f, spd$:
\begin{equation}\label{denominator-zero}
    \omega+\mu-E_f-\hbox{Re}\Sigma(\vec{k},\omega)=\,
    V^{2}/\left( \omega+ \mu- \varepsilon_{\vec{k}}\right)
\end{equation}
\begin{figure}[!tb]
\begin{center}
\includegraphics[width=8.0cm]{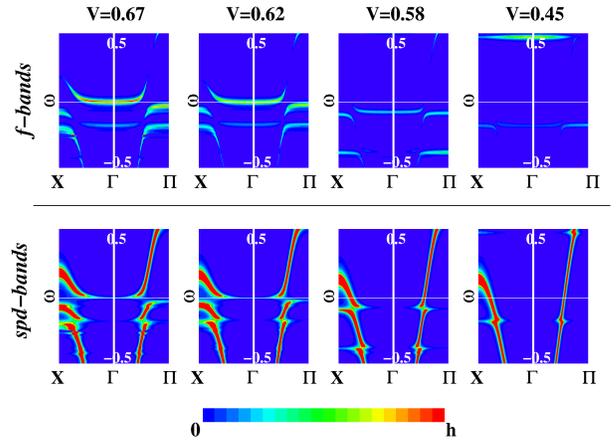}
\caption{Evolution across the transition point ($V^{*}\sim 0.58$)
of the $f$- (top row) and $spd$- (bottom row)
electron spectral functions along the path X=$(\pi,0,\pi)\to \Gamma=
(0,0,0)\to \Pi= (\pi,\pi,\pi)$ in momentum space.
The color scale
(bottom legend) is $h=1.0 [2.0]$ for $f$[$spd$]-electrons.}
\label{bands}
\end{center}
\end{figure}
The spectral-weight contribution to the electronic bands coming
from $f$ and $spd$ electrons are however very different. In
Fig. \ref{bands} we show the $f$ (top row) and $spd$ (bottom row)
spectral functions $-\frac{1}{\pi}\hbox{Im}G_{f[spd]}(\vec{k},\omega)$
along the path $X=(\pi,0,\pi)\to \Gamma=(0,0,0)\to
\Pi=(\pi,\pi,\pi)$ of momentum space, for varying
hybridization-parameter $V$ (from left to right).
For $V=0.67> V^{*}$ the band crossing
the Fermi level has predominantly $f$-character at low energy and
a strongly renormalized effective mass. In approaching the
transition point the $f$ electron contribution quickly reduces
until disappearing completely from the Fermi level for $V \sim
0.58 = V^{*}$. In addition, beyond the transition the $f$ band
shifts to negative energies. In describing the localization of
the $f$ electrons therefore, the double effect of suppression and
translation of the $f$ band has to be taken into account. Our
result is a prediction that can be observed in photo-emission
experiments. At the same time, in crossing the transition,
the effective mass of the $spd$ electrons
is reduced to the free value.

Recent studies\cite{stanescu06} have shown that insights into
quantum phase transition phenomena can be attained by studying
not only the Fermi surface (i.e. poles of the Green's function),
but also surfaces of zeroes in the Green's functions (i.e. poles
of the self-energy). We first remark that in our model there is
always a $G_{f}=0$ surface in momentum space, corresponding to
the free conduction electron Fermi surface FS0 (given by
$\varepsilon_{\vec{k}}- \mu= 0$). Further surfaces of zeroes ZS
in $G_{f}$ can appear in $\vec{k}$-space if there are $\vec{k}$
points for which $\Sigma(\vec{k},0) \rightarrow \infty$. In this
case we observe that $G_{spd}$ reduces to the free
conduction electron Green's function (Eq.~\ref{Gf}).
We show that this latter
phenomenon indeed takes place in approaching the transition point
$V^{*}$.
\begin{figure}[!tb]
\begin{center}
\includegraphics[width=8.0cm,angle=-0]{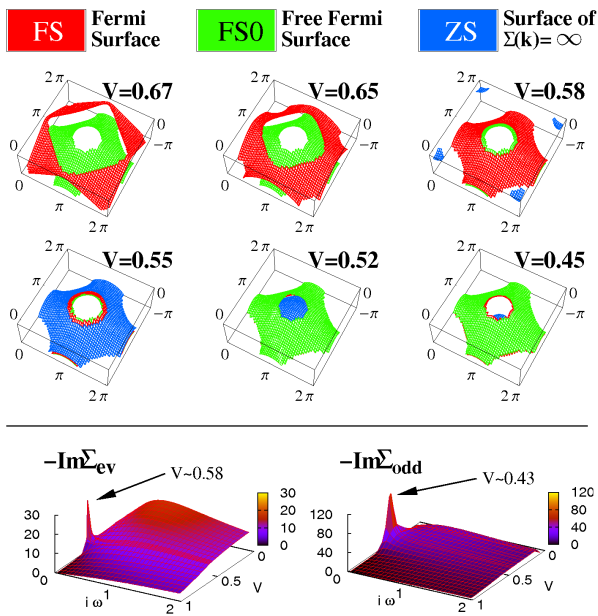}
\caption{Evolution of the Fermi Surface (FS, red) and the surface
of diverging self-energy (ZS, blue) across
the transition point $V^*$. An important role
is played by the free
conduction electron Fermi Surface (FS0, green).
The Mott character of
the transition is marked by the divergence of the
self-energy, detected by the cluster
self-energy eigenvalues $\Sigma_{ev}$ at $V\sim 0.58$ and
$\Sigma_{odd}$ at $V\sim 0.43$, as shown in the $V-i\omega_{n}$ space
(bottom panel).} \label{Fs-zeroes-free}
\end{center}
\end{figure}
In Fig. \ref{Fs-zeroes-free} we present the Fermi Surface
FS (determined by $\omega\rightarrow 0$
in Eq.~\ref{denominator-zero}),
FS0 and ZS
for different values of the
hybridizing coupling $V$ across the transition point. For
convenience sake, only the lower half of the 3-dimensional
Brillouin zone is shown.
In the heavy-fermion phase $V=0.67
> V^*$, only FS and FS0 are visible at
$\omega=0$ and far apart in momentum space.
In this case FS0 is not relevant for the
low-energy physics of the system.
As soon as $V\lesssim V^{*}$, however, a small
ZS appears around the point
$\vec{k}=(0,0,0)$ 
which pushes FS to collapse
onto the free FS0.
Since at FS0 $G_{f}\to 0$, this effect originates the
strong suppression and disappearance of the $f$ spectral weight
at the Fermi level (see Fig. \ref{bands}).
The appearance of ZS can be already seen in the
cluster quantities, which are displayed in the $V-i\omega_n$
space in the bottom of Fig. \ref{Fs-zeroes-free}.
At $V=V^{*}$ a divergence takes place for
$\omega_n\to0$\cite{deLeo07} in the even eigenvalue of the
cluster self-energy $\Sigma_{ev}$ (left panel), which, via the
periodization procedure, corresponds to the lattice
self-energy at $\vec{k}=(0,0,0)$. By
further reducing $V$ below $V^{*}$, ZS travels from $\vec{k}=(0,0,0)$ to
$\vec{k}=(\pi,\pi,-\pi)$,
where the divergence appears in
$\Sigma_{odd}$ for $V=0.43$ (right panel).
The position in $\vec{k}$-space of the FS remains unchanged for
$V<0.58$, numerically overlapping with FS0.
This indicates that at the Fermi level
$G_{f}=0$, i.e. the $f$ electrons
remain in a Mott state, and $G_{spd}$
reduces to the free Green's function.

The appearance of a divergent self-energy
proves that mottness is the
physical mechanism governing the localization of $f$-electrons.
In an
OSMT not all orbitals undergo a localization. In the metallic phase all the
orbitals participate in determining the Fermi volume, but, after the transition
took place, some ``selected'' orbitals do not contribute to the Luttinger
counting anymore. Across this transition a change in the compressibility of the
system is expected as localized orbitals become incompressible.
This is observed in the actinide series,
where the Mott transition can be driven e.g. by pressure\cite{Am}.
A Mott transition is also characterized by a significant rearrangement of the
electronic structure, since there is a transfer of spectral weight from low to
high energies. In our case the spectral weight is not entirely transferred from
the Fermi level to the Hubbard bands, but to an intermediate energy scale
giving rise to a pseudogap. The resulting modifications of the
quasiparticle dispersion can be understood in terms of divergence of the self-energy
similarly e.g. to the pseudogap of cuprates (see ref.\cite{stanescu06}).
Hence, in general, both a Fermi volume change and a significant rearrangement
of the bands are expected when a material undergoes a Mott transition.
Experimentally this would be detected by jumps in the Hall
coefficient\cite{paschen04} and in the de Haas-van Alphen frequencies\cite{shishido05}.
Measuring the phonon dispersions as a function of temperature
is another powerful probe of the orbitally selective Mott transition, as
suggested in Ref.\cite{falter}.
Furthermore the analysis of the 2IAM underlying our
self-consistent solution\cite{deLeo07} suggests other
experimental predictions. Close to the transition
the particle-hole symmetry breaking in the lattice model
generates a leading irrelevant operator which
is forbidden in the symmetric case and that causes a $\log T$
divergence in the specific heat coefficient at high temperature
\cite{2IK}, together with a $\log T$ divergence in the staggered
spin susceptibility and in the pairing susceptibility.
On the other hand, the formation of singlet correlations on the
energy scale of the pseudogap in the Mott insulating phase
implies a depression of the uniform spin susceptibility at low
temperature which has been observed for example in $CeRhIn_5$ but
not in $CeCoIn_5$\cite{zapf02,paglione_private}.
Since these effects originate from the competition of Kondo screening
and RKKY interaction, we expect them to take place below a temperature
comparable to (the largest of) these two energy scales (roughly
$10^{-3}$ the bandwidth, corresponding
approximately to 10 K in $CeRh[Co]In_5$).
This can be
understood considering that $CeRhIn_5$, unlike $CeCoIn_5$, is
antiferromagnetic at zero temperature and hence lies on the Mott
selective side of the transition. Optical conductivity (not
shown) displays a clear hybridization gap in the
delocalized phase which is absent in the phase where the $f$
electrons are localized. This is consistent with the experimental
assignment of $CeRhIn_5$ to the localized side and $CeCoIn_5$
to the itinerant side of the transition~\cite{burch07}.

To summarize, we have discussed the evolution of the momentum resolved
spectra of the periodic Anderson model across the quantum
phase transition, showing
its orbital selective Mott character. We have
described how the electronic structure undergoes dramatic
reconstruction: at the transition the $f$ spectral weight is
completely suppressed at the chemical potential and
a new energy scale emerges in the form of an $f$-electron pseudogap.
Within this latter energy range, the $spd$ band reduces to the free
band. $f$ and $spd$ electrons become totally decoupled at
low-energy while retaining a finite hybridization at higher
energies. We have finally shown that the concept of a surface of
diverging self-energy is useful for the understanding of
this phenomenon.

It is important to stress that the orbital selective Mott phase studied here is
not a stable phase at $T=0$ because the $f$ electrons order magnetically
as soon as they decouple at low energy from the conduction band\cite{deLeo07}.
In our calculations the magnetic ordering originates from an instability of the
orbital selective Mott state and not from an instability of the itinerant
paramagnetic heavy-fermion state. This supports the interpretation of the
magnetic transition as a byproduct of the OSMT (see also Ref.\cite{lanata}).
Our CDMFT study improves previous DMFT studies\cite{demedici05},
where the local character of the theory forbids the $T=0$ OSMT,
and previous slave boson studies
\cite{pepin07,Paul07,Coleman05,Senthil04}, where a finite
bandwidth in the $f$ electrons must be introduced in order to
have an exchange mechanism that is not killed by the vanishing of
the effective hybridization. In our case
such a term is not needed,
because retaining the full frequency dependence of
the self-energy allows to have a vanishing effective
hybridization at the Fermi level,
but at the same time an exchange mechanism
generated by the non-vanishing hybridization at finite frequency.

We acknowledge useful discussions with M.~Fabrizio, A.~Georges, C.~Pepin, 
J.~Paglione and I.~Paul. This work was supported by NSF grant DMR 0528969
and ICAM.


\begin{thebibliography}{99}

\bibitem{rmp07-heavyfermions}
H.v. L\"{o}hneysen, A. Rosch, M. Vojta, and P.
W\"{o}lfle, Rev. Mod. Phys. {\bf 79}, 1015 (2007);
P. Gegenwart, Q. Si, and F. Steglich, Nature Physics {\bf 4}, 186 (2008).

\bibitem{johansson}
B. Johansson, Phil. Mag. {\bf 30}, 469 (1974);
J.W. Allen, and R.M. Martin, Phys. Rev. Lett. {\bf 49}, 1106 (1982);
M. Lavagna, C. Lacroix, and M. Cyrot, Phys. Lett. A {\bf 90}, 210 (1982);
H. Watanabe, and M. Ogata, Phys. Rev. Lett. {\bf 99}, 136401 (2007).

\bibitem{lanata}
N. Lanat\`a, P. Barone, and M. Fabrizio, Phys. Rev. B {\bf 78}, 155127 (2008).

\bibitem{demedici05}
L. de' Medici, A. Georges, G. Kotliar, and S. Biermann,
Phys. Rev. Lett. {\bf 95}, 066402 (2005).

\bibitem{pepin07}
C. P\'epin, Phys. Rev. Lett. {\bf 98}, 206401 (2007).

\bibitem{rmp96}
A. Georges, G. Kotliar, W. Krauth, and M. J. Rozenberg, Rev. of
Mod. Phys. {\bf 68}, 13 (1996).

\bibitem{dmft-heavy}
K. Held, A. K. McMahan, and R. T. Scalettar,
Phys. Rev. Lett. {\bf 87}, 276404 (2001);
M.B. Z\"{o}lfl {\em et al.},
Phys. Rev. Lett. {\bf 87}, 276403 (2001);
K. Haule, V. Oudovenko, S.Y. Savrasov, and G. Kotliar,
Phys. Rev. Lett. {\bf 94}, 036401 (2005);

\bibitem{Paul07}
I. Paul, C. P\'epin, and M. R. Norman, Phys. Rev. Lett. {\bf 98},
026402 (2007).

\bibitem{pepin08}
C. P\'epin, Phys. Rev. B {\bf 77}, 245129 (2008).

\bibitem{kotliar01}
G. Kotliar, S.Y. Savrasov, G. P\'alsson, and G. Biroli, Phys.
Rev. Lett. {\bf 87}, 186401 (2001).

\bibitem{deLeo07}
L. De Leo, M. Civelli, and G. Kotliar, Phys. Rev. B {\bf 77},
075107 (2008).

\bibitem{stanescu07}
F. H. L. Essler, and A. M. Tsvelik, Phys. Rev. B {\bf 65}, 115117 (2002);
I. E. Dzyaloshinski, Phys. Rev. B {\bf 68}, 085113 (2003);
T. D. Stanescu, P. W. Phillips, and Ting-Pong Choy, Phys. Rev. B {\bf 75},
104503 (2007).

\bibitem{caffarel94}
M. Caffarel, and W. Krauth, Phys. Rev. Lett. {\bf 72}, 1545 (1994).

\bibitem{rmp06}
G. Kotliar {\em et al.},
Rev. of Mod. Phys. {\bf 78}, 000865 (2006).

\bibitem{stanescu06}
T.D. Stanescu, and G. Kotliar, Phys. Rev. B {\bf 74}, 125110
(2006); T.D. Stanescu, M. Civelli, and G. Kotliar, An. Phys. {\bf
321}, 1682 (2006).

\bibitem{Continentino}
M.~A. Continentino, Braz. J. Phys. {\bf 35} 197 (2005).

\bibitem{Am}
S. Heathman {\em et al.},
Phys. Rev. Lett. {\bf 85}, 2961 (2000).

\bibitem{paschen04}
S. Paschen {\em et al.},
Nature  {\bf 432}, 881 (2004).

\bibitem{shishido05}
H. Shishido, R. Settai, H. Harima, and Y. O-nuki, J. Phys. Soc.
Jpn. {\bf 74}, 1103 (2005).

\bibitem{falter}
C. Falter, T. Bauer, and F. Schnetg\"oke, Phys. Rev. B {\bf 73}, 224502 (2006).

\bibitem{2IK}
I. Affleck, A.W.W. Ludwig, and B. A. Jones, Phys. Rev. B {\bf 52}, 9528 (1995).

\bibitem{zapf02}
V. S. Zapf {\em et al.}, Phys. Rev. B {\bf 65}, 014506 (2001).

\bibitem{paglione_private}
J. Paglione, private communication.

\bibitem{burch07}
K. S. Burch {\em et al.}, Phys. Rev. B {\bf 75}, 054523 (2007).

\bibitem{Coleman05}
P. Coleman, J. B. Marston, and A. J. Schofield, Phys. Rev. B. {\bf 72},
245111 (2005).

\bibitem{Senthil04}
T. Senthil, M. Vojta, and S. Sachdev, Phys. Rev. B {\bf 69},
035111 (2004).



\end{thebibliography}
\end{document}